# Inkjet printed intelligent reflecting surface (IRS) for indoor applications

Kairi Takimoto[1], Kazutomo Nakamura[1], Peter Njogu[1], Kota Suzuki[1], Masato Sugimoto[1], Ashif Fathnan[1], Takashi Kondo[2], Masayuki Mori[2], Daisuke Anzai[1], and Hiroki Wakatsuchi[1]

[1] Department of Engineering, Nagoya Institute of Technology, Gokisocho, Showa Ward, Nagoya, Aichi 466-0061, Japan
[2] GOCCO Inc., 2-26 Hommachi, Ogaki, Gifu 503-0885, Japan

Email: wakatsuchi.hiroki@nitech.ac.jp

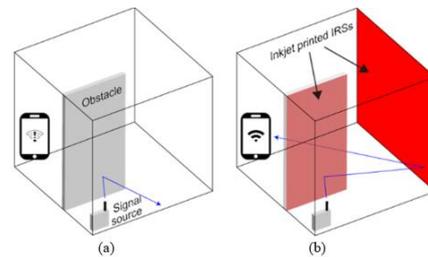

**Fig 1** *Illustrative indoor NLOS communication scenario (a) without (b) with IRSs.*

Abstract. A passive, low-cost, paper-based intelligent reflecting surface (IRS) is designed to reflect a signal in a desired direction to overcome non-line-of-sight scenarios in indoor environments. The IRS is fabricated using conductive silver ink printed on paper with a specific nanoparticle arrangement, yielding a cost-effective paper-based IRS that can easily be mass-produced. Full-wave numerical simulation results were consistent with measurement results, demonstrating the IRS's ability to reflect incident waves into a desired nonspecular direction based on the inkjet-printed design and materials.

*Introduction:* In indoor environments, wireless signal dead spots i.e., non-line of sight (NLOS) conditions, occur due to the physical presence of obstacles, e.g. walls, buildings or large objects, that block or absorb the radio-frequency (RF) signals. These obstacles partially or fully block the direct line of sight between the transmitting and receiving antennas, causing signal attenuation or complete signal loss, as illustrated in Figure 1(a). To overcome this challenge, various methods have been employed, such as using signal repeaters [1], distributed antenna systems [2], mesh networks [3], frequency selective surfaces [4,5] and dipole scatterers [6]. More recently, the concept of intelligent reflecting surfaces (IRSs) has emerged as a potential solution to overcome NLOS scenarios utilizing metasurfaces that efficiently reflect incident waves into arbitrary nonspecular angles [7–9]. By placing these surfaces strategically in an indoor environment, they can redirect or reflect RF signals around obstacles, effectively reducing or eliminating the shadowed areas. Reconfigurable IRS options, both active [10,11] and passive [12,13], have also been suggested to control beamforming patterns by incorporating electronic circuits within the surface. However, considering that existing approaches are often costly and require a printed circuit board (PCB) and many electronic components, including external energy source, a low-cost yet effective IRS solution is more desirable, particularly, in indoor environments. Based on the preceding discussion, an affordable IRS solution created through inkjet printing is presented here. Compared to active PCB-based IRSs, the proposed IRS does not require a direct-current (DC) supply, making installing it easier.

The application of inkjet printing as fabrication method for electronic components, such as antennas [14] and frequency-selective surfaces [5], on paper substrates, has been adopted in various studies to create efficient and cost-effective printed devices. Here, we introduce the design, fabrication, and experimental demonstration of an inkjet-printed paper-based IRS, which can improve indoor NLOS signal propagation. Its array with size-varying patches guides the signal in a desired direction without requiring embedded electronic circuits. Using an ordinary inkjet printer, the proposed application is inexpensive and easy to fabricate on readily available paper substrate. When placed on walls or household furniture, it can reflect waves anomalously and avoid complete signal blockage by indoor obstacles (see Figure 1(b)). The proposed IRS offers potential as a cost-effective solution for indoor communication, with the prospect of large-scale production due to its low fabrication expenses.

*Materials and methods:* The IRS was designed to operate at 5 GHz with ten unit cells per supercell with a period $P$ of 120 mm. Based on results from the ANSYS HFSS modeller, the unit cell of Figure 2(a), whose dimension are shown in Table 1, was designed and simulated on a paper substrate with a relative permittivity of 2 [15] and tan(δ)=0.05. The gap between the metasurface patches and the copper ground was optimized by including a 1.0 mm thick medium-density-fibreboard (MDF) piece with a relative permittivity of approximately 2.5 [16] to enhance the reflection process. Figure 2(b) is the reflective properties of the unit cell while Figure 2(c), is the 240 mm × 360 mm supercell of dimensions shown in Table 2.

A metasurface with abrupt-changing material properties exhibits anomalous manipulation capabilities of an incident wave [17,18] in accordance with the generalized Snell's law with an extra parallel wave vector due to the resultant radiation phase gradient created on the metasurface. A field pattern is reflected by the inhomogeneous metasurface when a normally incident signal illuminates the metasurface. This is due to the abrupt-varying material properties of the metasurface. Wavefront shaping controls the direction of reflection by designing the gradient of the reflection phase through the supercell array with size-varying patches, Figure 3(a). The size of each patch determines the reflection angle. As each of unit cells radiates with a different phase, $\Phi_y$, the interaction between the incoming signal and the different unit cells creates a new wavefront as defined by the slanting blue line of Figure 3(a), and generates a phase gradient, $\frac{d\Phi_y}{dx}$. If the phase gradient is constant, i.e., $\frac{d\Phi_y}{dx} = \varrho$, then it can be demonstrated that the reflected wave is a plane wave carrying parallel wave vector $k_x^r = \varrho$, in accordance with the generalized Snell's law derived in [17]. Thus, $\varrho = \frac{2\pi}{P}$, $k_0 = \frac{2\pi}{\lambda}$, where $P$ = 120 mm and $k_0$ is the wave vector at wavelength $\lambda$. Thus, the parallel wavevector is:

$$k_x^r = 0.5 k_0 \quad (1)$$

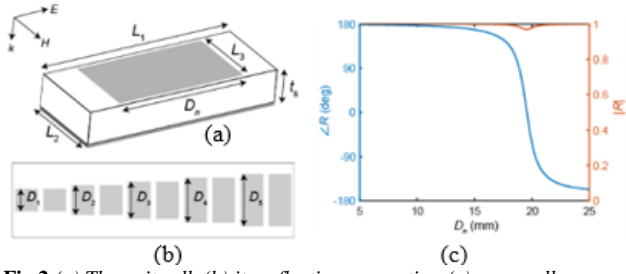

**Fig 2** *(a) The unit cell, (b) its reflection properties, (c) supercell.*

*Table 1. Unit cell dimensions.*

| Dimension | $L_1$ | $L_2$ | $L_3$ | $D_n$ | $t_s$ |
|---|---|---|---|---|---|
| Value (mm) | 30 | 12 | 11 | variable | 1.1 |

*Table 2. The supercell dimensions.*

| Dimension | $D_1$ | $D_2$ | $D_3$ | $D_4$ | $D_5$ | $P$ |
|---|---|---|---|---|---|---|
| Value (mm) | 16.4 | 19.1 | 19.6 | 20.1 | 21.3 | 120 |

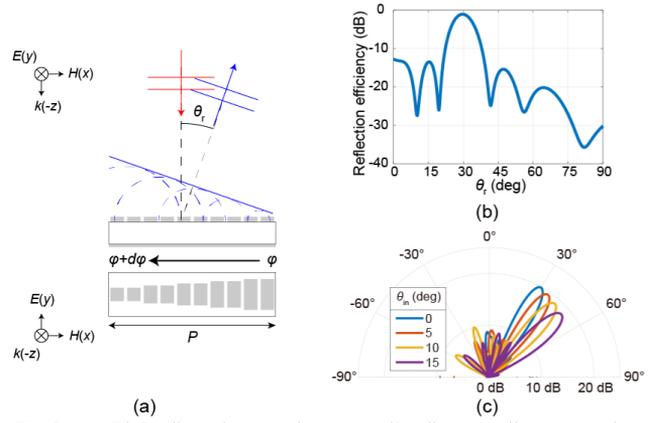

**Fig 3** *(a) The reflected wave, (b) supercell reflection efficiency and (c) simulated far-field reflection profile of the metasurface.*

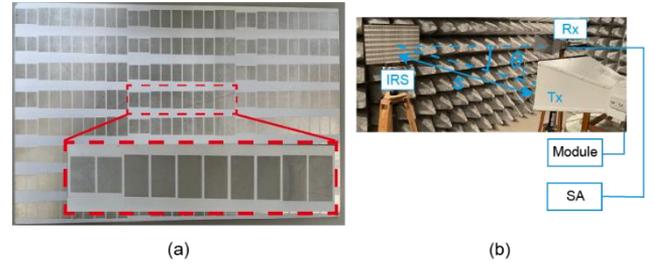

**Fig 4** *(a) The IRS sample and (b) the measurement setup.*

As per the generalized Snell's law, the IRS thus reflects a normally incident wave along angle $\theta_r$ as:

$$\theta_r = \arcsin(0.5) \approx 30° \quad (2)$$

For the oblique incidence case [18], the reflection angle $\theta_r$ is given by:

$$\theta_r = \arcsin(\sin\theta_i) + \frac{\varrho}{k_0}) \quad (3)$$

*Results:* Reflection efficiency of the supercell is shown in Figure 3(b) as a comparison of its reflection with that of PEC. Figure 3(c) depicts the metasurface's simulated far field anomalous reflection profiles in polar form for both normal and oblique angles incidence. It shows the peak directivity, i.e., the maximum ratio of the IRS's radiation strength in the given direction to radiation strength averaged over all directions. They are consistent with the analytical results of (2) and (3). The metasurface anomalously reflects a normally incident wave at 30°, while oblique waves at 5°, 10° and 15° incidence are reflected at 36°, 42° and 49°, respectively. The IRS was printed on a paper substrate using conductive silver ink cartridge and an inkjet printer. An MDF material was placed between the paper substrate and a copper ground plane. Figure 4(a) depicts the IRS metasurface sample and shows a supercell surrounded by dashed red lines. Figure 4(b) shows the measurement setup in the anechoic chamber.

A transmitter (Tx) was set normal to the metasurface at distance $s$ = 1.5 m, while a receiver (Rx) angle was set to 30°. The reflected signal was measured using a spectrum analyser (Tektronix, RSA306B). Tx was connected to a commercial Wi-Fi module (Micro research, MR-GM3) that generates a 14-dBm signal and supports the 5-GHz band wireless LAN communication standard (IEEE802.11ac/n/a) [19].

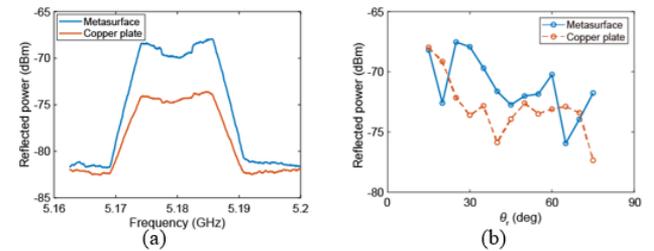

**Fig 5** *Reflected power at (a) $\theta_r$ = 30° and (b) various angles.*

Figure 5 shows the measured reflected power in the frequency domain from 5.16 to 5.20 GHz. An anomalously reflected signal was detected at 30° for the normally incident signal, Figure 5(a), consistent with the simulation results (Figure 3(c)). It also shows a stronger reflected signal from the metasurface compared to a metallic copper plate at the same angle shows. This behaviour is replicated in Figure 5(b) for oblique angles of incidence. This is because the reflection phase was inhomogeneously designed over the metasurface, so the reflection wavefront

2                              ELECTRONICS LETTERS wileyonlinelibrary.com/iet-el

was re-directed to propagate at 30°. To dynamically manipulate a wireless channel in a time-varying environment, multiple IRSs with different phase gradients could be used. Compared to existing wave controlling metasurfaces [20, 21], the proposed IRS is simple, low-cost and easy to deploy.

*Conclusion:* An inkjet-printed paper-based IRS that manipulates a reflected wave is presented here. The paper-based reflection scheme implemented by the proposed IRS produces a new wavefront when waves are reflected from the different resonators with a carefully designed reflection phase, thereby re-directing the signal at an anomalous angle. The proposed IRS offers a potential solution to NLOS communication, which is enabled by the anomalous reflection of the signal with an inexpensive paper-based metasurface. The resulting surface reflects an incident signal with a phase gradient profile over a $2\pi$ period. The low production cost of the device supports its potential for mass production. This work can be extended by incorporating nonlinear circuits and enabling waveform selectivity, which offers an additional degree of freedom to control RF signals [22]. Such advancements would enable new designs capable of, e.g., pulse-width-based redirection of a reflected signal [12] and preferential extraction of pulsed waveforms from a complex superimposed input signal [23]. These would add further extensibility to the proposed electromagnetic environment modification tool.

*Acknowledgments:* This work was supported in part by the National Institute of Information and Communications Technology (NICT), Japan under the commissioned research No. 06201, the Japan Science and Technology Agency (JST) under the Precursory Research for Embryonic Science and Technology (PRESTO) No. JPMJPR193A and under the Fusion Oriented Research for Disruptive Science and Technology (FOREST) and KAKENHI grants from the Japan Society for the Promotion of Science (JSPS) (Nos. 21H01324 and 22F22359).